# Impact on Mobility and Environmental data of COVID-19 Lockdown on Florence Area

(a DISIT lab Data Report for COVID-19, on Mobility and Environment Focus)

Paolo Nesi, paolo.nesi@unifi.it, Date: 07-05-2020, Version 0.2, Ref:

C. Badii, P. Bellini, S. Bilotta, D. Bologna, D. Cenni, A. Difino, A. Ipsaro Palesi,
N. Mitolo, P. Nesi, G. Pantaleo, I. Paoli, M. Paolucci, M. Soderi

Https://www.disit.org, https://www.snap4city.org

**Abstract**: According to the changed operative conditions due to lockdown and successive reopening a number of facts can be analysed. The main effects have been detected on: *mobility, environment, social media and people flows.* While in this first report only mobility, transport and environment are reported. The analysis performed identified a strong reduction of the mobility and transport activities, and in the pollutants. The mobility reduction has been assessed to be quite coherent with respect to what has been described by Google Global mobility report. On the other hand, in this paper a number of additional aspects have been put in evidence providing detailed aspects on mobility and parking that allowed us to better analyze the impact of the reopening on an eventual revamping of the infection. To this end, the collected data from the field have been compared from those of google and some considerations with respect to the Imperial college Report 20 have been derived. For the pollutant aspects, a relevant reduction on most of them has been measured and rationales are reported.

## 1. Introduction

Italy has been one of the first countries in Europe to experience the crisis provoked by the so-called infection COVID-19. To this end, the Italian govern implemented a number of Non-Pharmaceutical Intervention (NPIs), aiming at full lockdown, previously in specific regions and then in the whole country acting on reducing mobility, social interaction on school and university, social distancing, banning social events, closing public events, services and restaurants, etc. In Tuscany, where DISIT Lab with its Snap4City infrastructure and service has the largest amount of data collected from several sources, the lock down has been progressively performed from the 5$^{th}$ to the 10$^{th}$ of March. Moreover, the lockdown has been removed the 4$^{th}$ of May, starting with a progressive reopen of production activities, still leaving close the social events, entertainments, restaurants, etc., some have had to reorganize themselves to provide take-away services instead on their internal services and rooms. Also the research activity restarted, but we as DISIT lab never closed, we remain for the whole duration of the lockdown in smart working modality as we are now, continuously supporting the Snap4City infrastructure and services, and developing according to the large number of research projects we had in place in the period.

**In this paper**, according to the above facts, we would like to put in evidence which kind of effects we have detected in the data collected and from those data what can be deduced so far about the above described period from beginning of March 2020 up to May 2020 also with respect to the previous weeks, months and year for the same variables and data collected, observed. The main effects have been detected on: *mobility, environment, social media and people flows, while this report is mainly focussed on mobility, transport and environmental aspects.* For these aspects different data collected and deductions can be provided. Therefore, for each of these domains and/or for each kind of data a separate discussion is presented in the following sections. Moreover, the domains and data may not refer to the same area since the Snap4City data and services are covering a large number of cities and regions as described in **Figure 1**. For this reason, before presenting the data and the results a short overview of two main facilities of DISIT Lab involved is reported: (1) Snap4City solution and areas, (2) Twitter Vigilance for monitoring social media (which in this cases has not been adopted for the assessment of mobility and environmental aspects) and that will be subjects of a different paper [8], [9], [10].

The paper is organized as follows. In section 2, an overview of Snap4City is reported. Section 3 describes the dashboard for collecting and processing data coming from the Italian Civil Protection regarding COVID-19, Section 4 describes the impact on mobility about the lockdown and possible deductions. Section 5 described the impact of lockdown on parking facilities and deductions. In Section 6, the impact of lockdown on environmental data and deductions. In Section 7 conclusions are drawn.

## 2. Snap4City overview

Snap4city (https://www.snap4city.org ) has been developed to provide many online tools and guidelines to involve all different kinds of organizations (e.g., Research Centers and Universities, small business, large industries, public administrations, and local governments) and citizens (e.g., city operators, resource operators, companies, tech providers, category Associations, corporations, research groups, advertisers, city users, community builders) [1-7]. Full training on Snap4City is accessible on Https://www.snap4city.org/577

Snap4City improves city services, security and safety by offering a sustainable solution for smart city and Living Lab, thus attracting industries and stakeholders. Snap4City is capable to keep under control the real time city evolution: reading sensors; computing and controlling key performance indicators, KPI; detecting unexpected evolutions; performing analytics; taking actions on strategies and alarms. Snap4City supports the city in the process of continuous innovation on services, infrastructures, with control and supervision, tools for business intelligence, predictions, anomaly detection, early warning, risk assessment, what-if analysis, also setting up strategies for increasing city resilience with respect to unexpected unknown events.

Thanks to knowledge base support, Snap4City provides flexible solutions to get immediate insights and deductions of the city status and evolution, exploiting ultimate artificial intelligence, data analytics and big data technologies, activating sentient solutions collecting, and exploiting heterogeneous data of any kind, from any data source (open and private; static, real time, event driven, streams, certified and personal). Snap4City solution provides a flexible method and solution to quickly create a large range of smart city applications exploiting heterogeneous data and enabling services for stakeholders by IOT/IOE, data analytics and big data technologies.

Snap4City applications may exploit multiple paradigms as data driven, stream and batch processing, putting co-creation tools in the hands of: (i) Smart Living Lab users and developers a plethora of solutions to develop applications without vendor lock-in nor technology lock-in, (ii) final users customizable / flexible mobile Apps and tools, (iii) city operators and decision makers specialized / sophisticated city dashboards and IOT/IOE applications for city status monitoring, control and decision support. Snap4City satisfies all the expected requirements of ENOL, EIP-CPP, Select4Cities challenge PCP and much more, and it is 100% open source, scalable, robust, respects user needs and privacy; provides MicroServices and easily replaceable tools; compliant with GDPR; provides a set of tools for knowledge and living lab management, and it is compliant with more than 70 protocols including end-to-end encrypted communication.

Snap4City is an official platform of FiWare, an official library of JS Foundation Node-RED, registered on E015, present on EOSC marketplace, and BeeSmartCity MarketPlace, etc. Snap4City obtained the 1st place award by Select4Cities partners and PCP (Antwerp, Copenhagen and Helsinki).

Snap4City provides services and data of several cities/Organizations as: *Firenze, Helsinki, Antwerp, Lonato del Garda, Santiago de Compostela, Pisa, Prato, Pistoia, Lucca, Arezzo, Grosseto, Livorno, Siena, Massa, Modena, Cagliari, Valencia, Pont du Gard, Dubrovnik, West Greek, Mostar; and from regions as Tuscany, Garda Lake,*

*Sardegna, Belgium, Finland, Emilia Romagna, Spain, etc*. Snap4City is open to your contributions, using Snap4City tools and contributing in improving them, adding more tools and features, etc. Please join the community on this portal and on GitHub/disit. https://github.com/disit

## 3. Dashboarding Covid-19 data on Snap4City

According to the data provided by the Civil Protection and accessible from https://github.com/pcm-dpc/COVID-19, we have gathered in automatic the data from that data sources which is updated once a day at 17-18:00. The acquired data have been partially processed for providing a tool for data intelligence with a main focus of comparing the general national trend with respect to the trends of Tuscany region (see Figure 1). https://www.snap4city.org/dashboardSmartCity/view/index.php?iddasboard=MjU2OQ== The web page on COVID-19 at Snap4City is https://www.snap4city.org/599 .

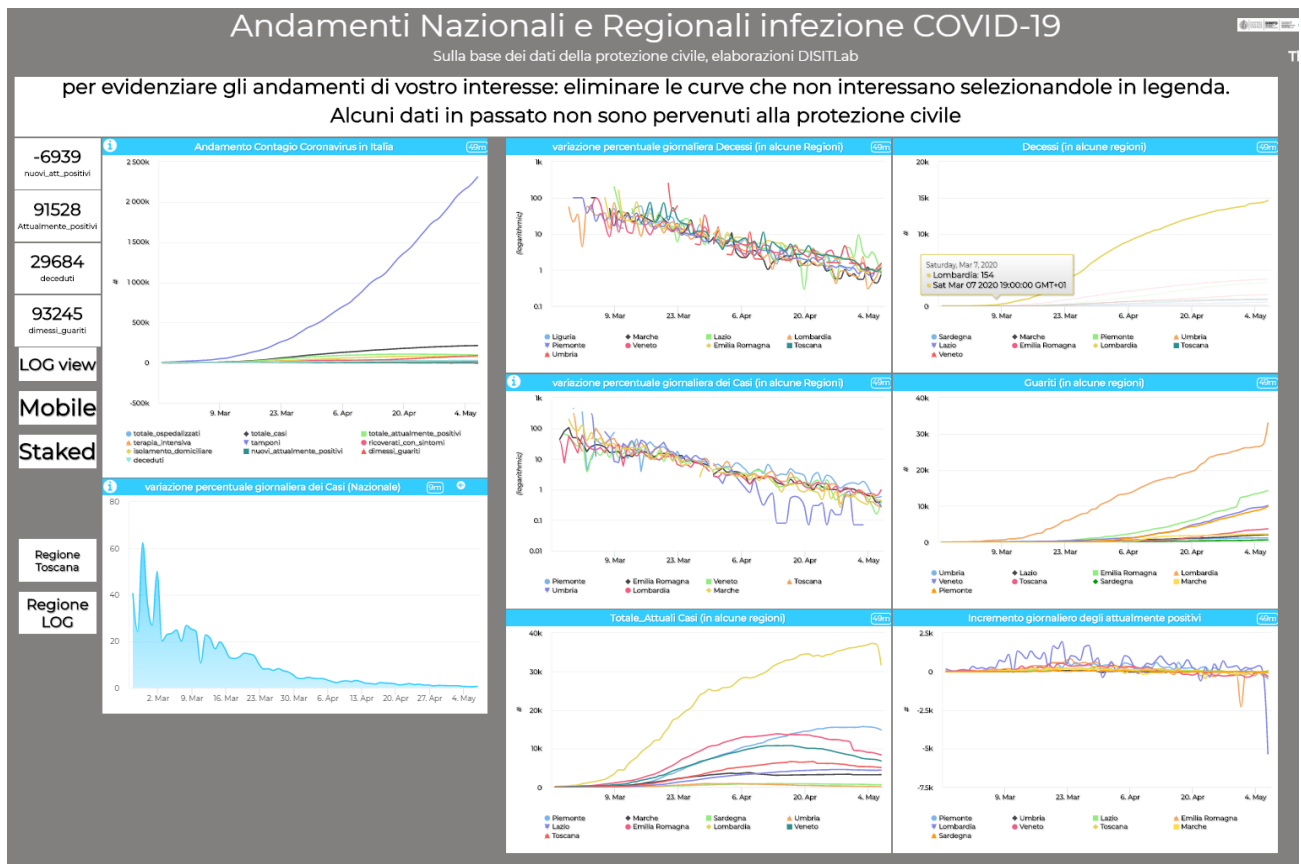

**Figure 1: Data visual analytics on the basis of the COVID-19 data coming from Italian Civil Protection**

The dashboard is in Italian language and includes: the different trend at national level, on the left bottom corner the % of variation of daily cases which is consolidated index of the infection, the % of variation for the deaths for the major regions, etc. It is also possible to pass at the regional view which can be reached directly with: https://www.snap4city.org/dashboardSmartCity/view/index.php?iddasboard=MjU3NQ== At the time of wiring of this paper, the most of the provinces in Tuscany has apparently reached a point close to maxim. It is even much more evident from the LOG view of the same graphs:

https://www.snap4city.org/dashboardSmartCity/view/index.php?iddasboard=MjU4MA==

This Dashboard reports the trends of cases for the 10 provinces of Tuscany Region, and the aligned trends at 100 cases of the curves for the most infected provinces in Italy with respect to those of Tuscany (on t he rights side of the Dashboard).  From this last picture it is evident that in Tuscany most of the provinces have a rate for cases which much lower with respect to the most infected provinces in Italy that are: Milano, Torino, Brescia and Bergamo.

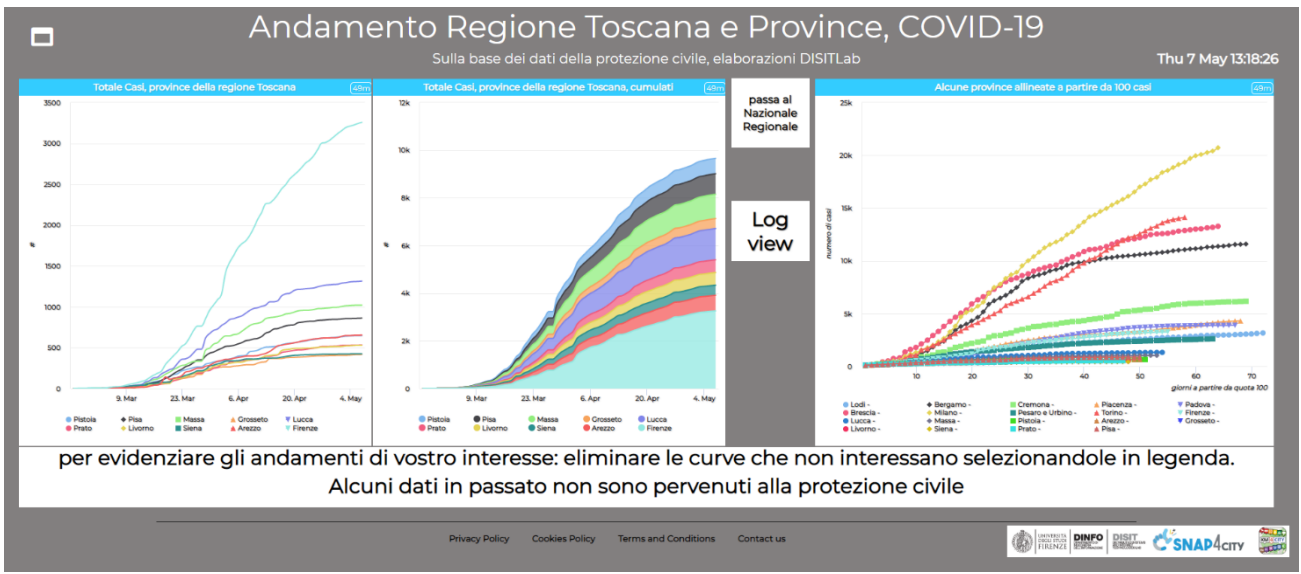

**Figure 2: Data visual analytics on the basis of the COVID-19 data coming from Italian Civil Protection**
https://www.snap4city.org/dashboardSmartCity/view/index.php?iddasboard=MjU3NQ==

## 4. Impact of lockdown on traffic data and deductions

In the context of mobility and transport relevant data sources that could be affected by the above described situation are **Traffic flow sensors (**which expressed the number of vehicles passing the road, and the density of vehicles per time slot, for example every 15 minutes) placed in the:

1. main roads of the city monitoring inflow and outflow of the city.
2. main entrance of the restricted traffic zone, RTZ
3. internal main roads of the city (not directly addressed in this report)
4. high speed road of the region (not directly addressed in this report)

For cases (1) and (2), the in/out flows are computed on the basis of the traffic flow sensors placed on the main roads connecting the city with the rest of countryside (see **Figure 3**). So that, those traffic sensors are describing the effective number of vehicles that enter/exit in/from the city every 10/15 minutes and thus allow to compute the total number of moving vehicles in the day. In Florence, in normal conditions we have an average of 290000 vehicles entering and exiting every day, in both direction and almost balancing the two flow. Where for *vehicles* we intend equivalent vehicles, which means that cars are counted 1, bus counted 2.5, motorbikes 0.5, etc.

Please note that, taking into account traffic flow sensors, it is possible to reconstruct the traffic flow in the other road segments of the city in which the flow sensors are not present as described in [11], [12]. See also: https://www.snap4city.org/dashboardSmartCity/view/index.php?iddasboard=MTc5NQ== for public Dashboard with traffic flow reconstruction in real time. The traffic flow reconstruction allows to actually understand which is the effective usage of the city roads without the need to have a dense network of expensive sensors.

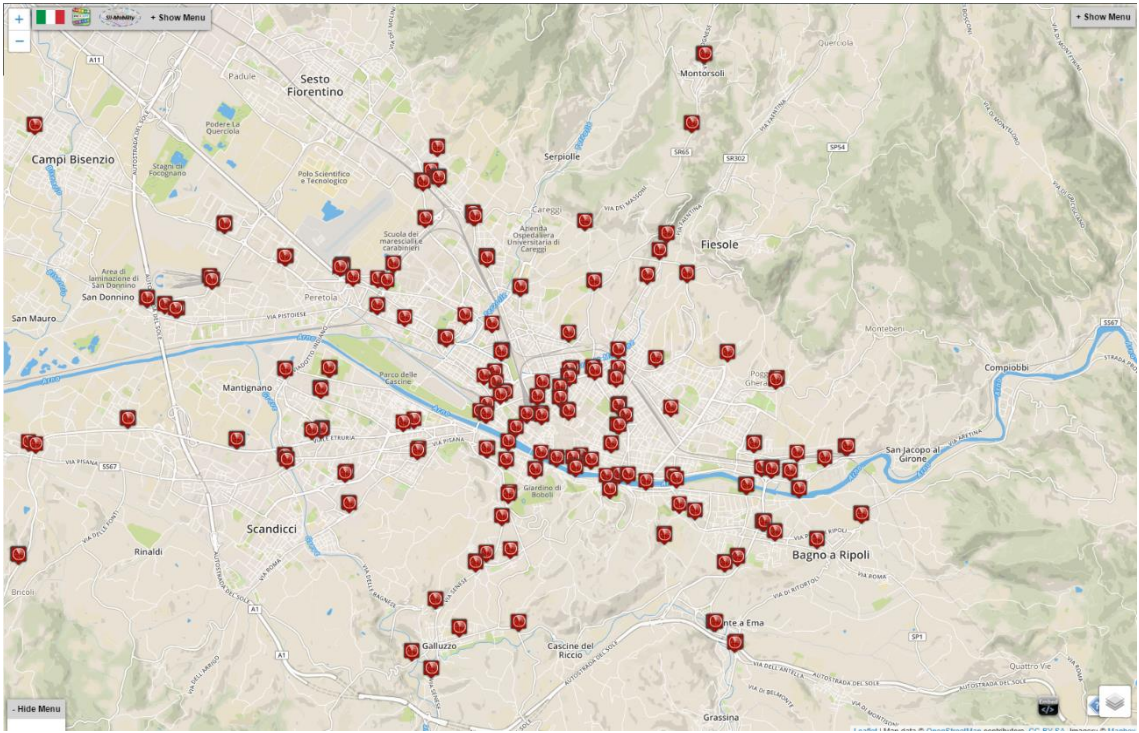

**Figure 3: placement of traffic flow sensors in the city, https://servicemap.snap4city.org/**

As regarding the **traffic flow**, the following **Figure 4** shows the impact of the lockdown on the inflow and outflow of *vehicles* in Florence (which is the point 1 of the above list). The first part of **Figure 4** Dashboard reports the trend over H24 for in/out flows of the city and RTZ inflow, compared with the trend of the previous day. In the second part of the dashboard, the trend of daily counting of vehicles entering (inflow) and exiting (outflow) from Florence city and its RTZ inflow, in the last 12 months are reported, and compared with the previous year data (in grey). Please note that the graphs present a weekly periodic trend (see also canyons due to the weekends), and some holes due to mistakes in the sensors network and communications.

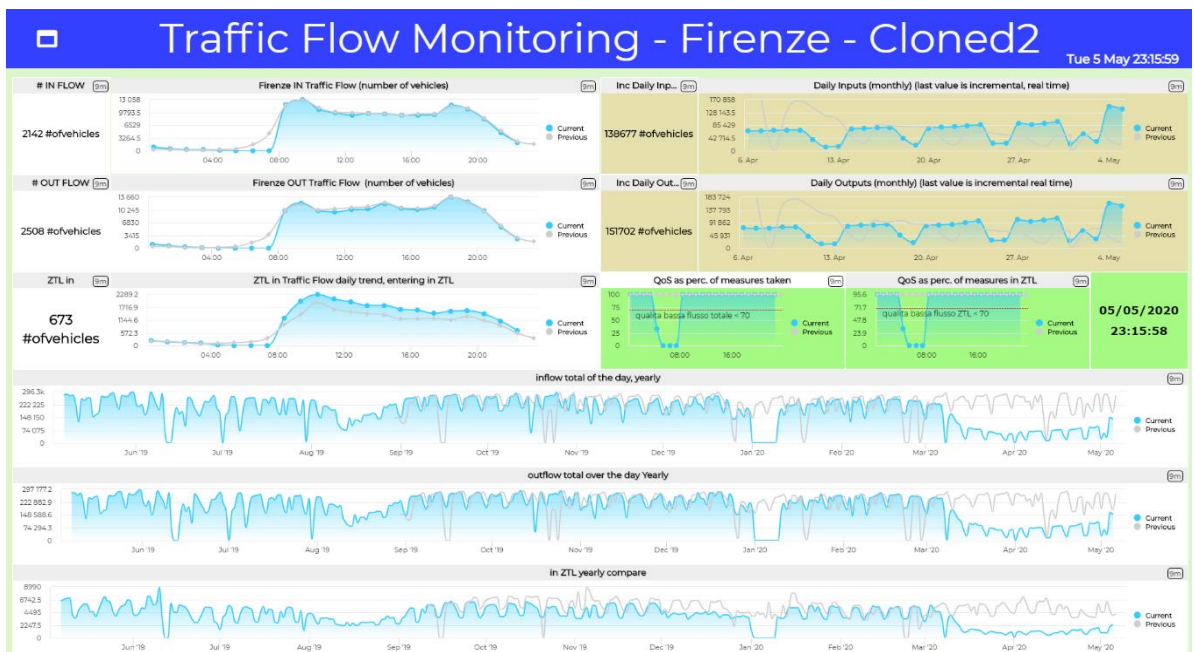

**Figure 4 – Traffic flow dashboard from Snap4City for Florence area (see detail in Figure 3), accessible from:**
https://www.snap4city.org/dashboardSmartCity/view/index.php?iddasboard=MjY1MQ==

In more detail, the following graphs report the trends of whole daily counting of vehicles entering (inflow) and exiting (outflow) from Florence in the period of January and May 2020 (in blue), with respect the corresponding values measured on the 2019 on the same dates (in grey). Please note that, the lockdown has been started effectively the 9th of March (a Monday), while in the first days of April the total flow traffic was reduced to the 18% (53000 with respect to the 284000). On the other hand, the 4th and 5th of May (the first and second days of the reopening) the reduction was still of the 52% (148.000 with respect to 284000). It can be similarly stated for the RTZ, where the reduction in the first days of April has been at the 20%, and at the reopen in May of 46%. This reduction does not take into account the movements of people that are going to move by working or biking and neither in the flow internal to the city, a part for the flow into the RTZ. On the other hand, a similar reduction has been also recorded taking into account all the other traffic flow sensors inside the city as in **Figure 3**, and along the high-speed road as FiPiLi in Tuscany.

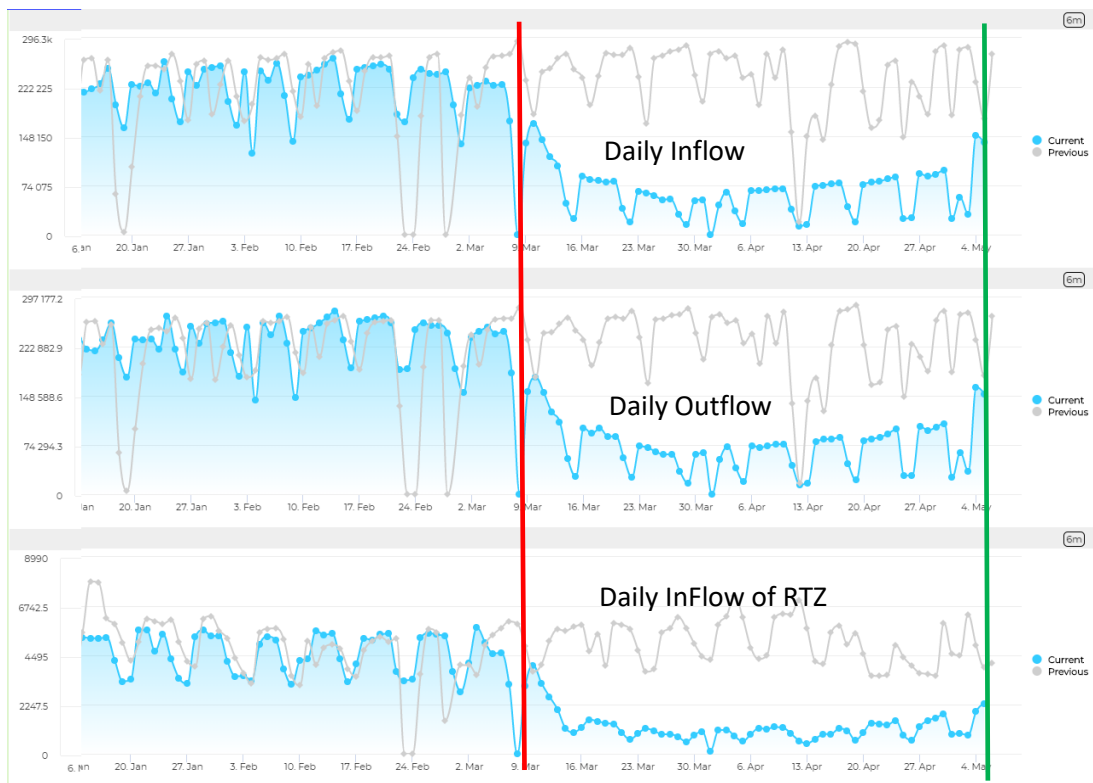

**Figure 5 Traffic flow dashboard from Snap4City for Florence area (red is the start of the lockdown, and green the restart), accessible from:** https://www.snap4city.org/dashboardSmartCity/view/index.php?iddasboard=MjY1MQ==

Similarly, from Google Global Mobility report [13], [14], it can be observed a reduction of mobility (in this case measured from the Google mobile Apps) of the people as reported in **Figure 6**. In this case, as reported by Google the reduction has been at the 30% regarding the baseline (so that -70%), at 75% for grocery & pharmacy, at 30% for Parking, at 33% for transit in stations, at 45% for working, while an increment of local residential movements have been recorded with at a 128% with respect to the previous conditions. Comparing the data of Google with respect to those collected from the actual sensors, it appear evident that the data of Google cannot be combined each other since the different trends are only relative reductions or increment with respect to the corresponding baseline that is unknown. So that, the actual reduction of mobility for working in terms of number of people is unknown. On the other hand, the reduction to the 30-25% of mobility in confirmed for baseline, parking and transit stations; which is coherent with more precise values estimated by sensors still in Florence area.

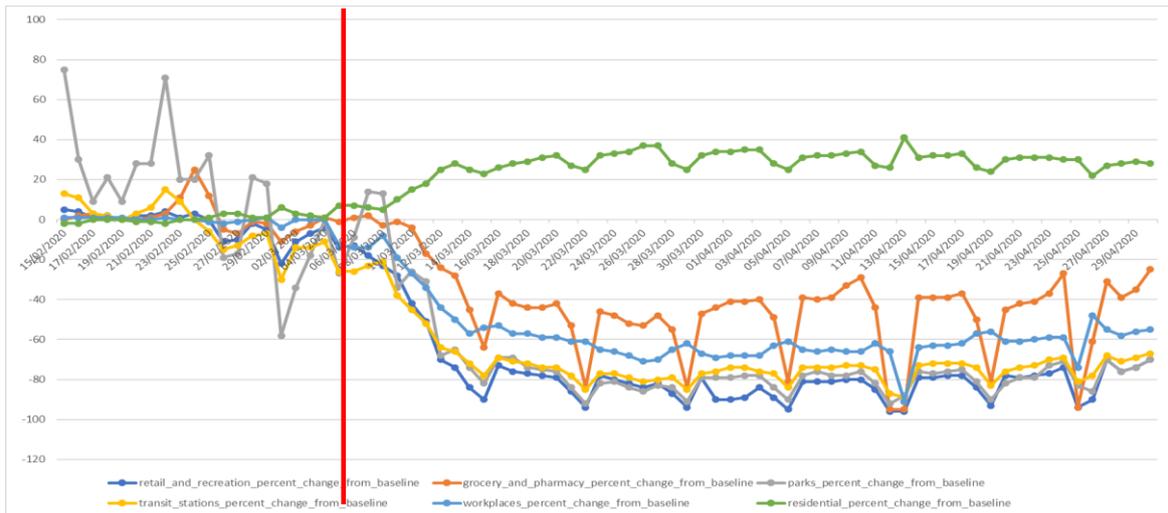

Figure 6: reduction of mobility from Google Global mobility data [13], [14]. (red line marks the start of the lockdown)

The remaining 20% of traffic flow during lock down has been mainly due to the mobility of mandatory services for example for the activities to guarantee the functionalities of hospital, supermarkets, pharmacies, public transportation for those workers, etc. In [15] (the Report 20 of Imperial College), the effect of the recovery of traffic and mobility of +20% and +40% from the 4$^{th}$ of May with respect to the lockdown conditions have been analysed. In [15], the model proposed puts in connection the increment of traffic with the R factor that should remain below 1 to control the epidemic. In **Figure 7**, the Simulation presented in [1] regarding Tuscany for different cases in which a different rate of increment of mobility is supposed.

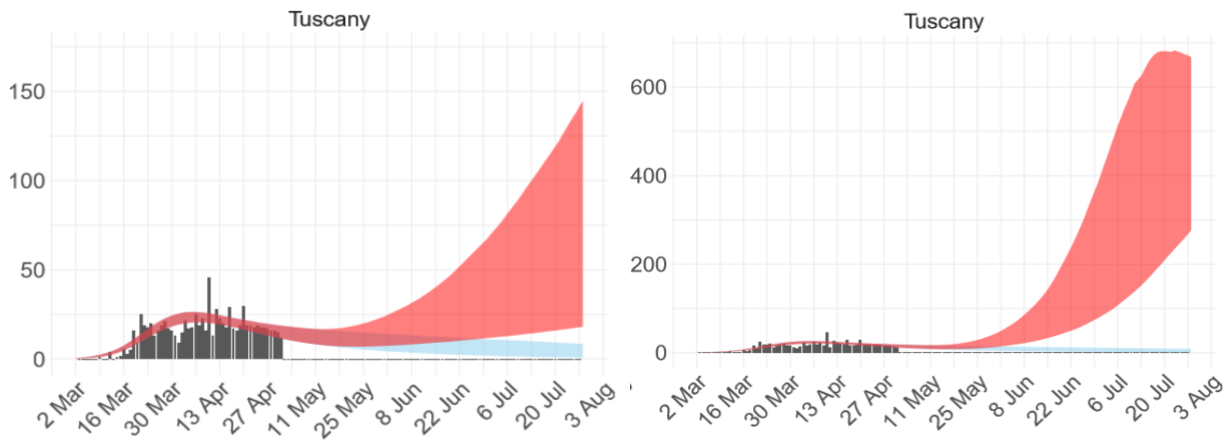

Figure 7: Simulation presented in [15] regarding Tuscany: (i) in red the trend of the deaths as confidence intervals over time in the cases of +20 (on the left), +40 (on the right) of mobility, (ii) in cyan the trend of the confidence values over time of the possible deaths in the cases of lock down continuation, and (iii) in black the officially registered number of deaths.

The effective impact of the registered increment of mobility in Florence is very hard to be assessed since the model is supposing that the returning of the mobility would be performed with the same human behaviour of before COVID-19. On the other hand, most of the population has understood to use the NPIs. Moreover, the model presented in [15] does not clarify which amount of mobility they have suppose to have during lockdown. If they have supposed the mobility very close to zero, the model in that sense has been optimistic. This means that the actual forecast of the number deaths with respect to the increment of traffic and mobility is very hard to be produced even with large bounds of confidence.

## 5. Impact of lockdown on parking facilities and deductions

In the context of mobility and transport relevant data sources that could be affected by the above described situation are **Parking status sensors** placed in the most relevant parking infrastructures (which allows to count the number of free parking slots over time) at services of:

- o Hospital areas
- o railways stations
- o justice area
- o markets and shopping centers
- o social and entertainment areas

In Florence, and Tuscany there is a number of infrastructure parking lots in which the number free spaces are counted directly at the entrance/exit. They are located in the whole region. On the other hand, we are focussing on this paper only on the major parking lots in Florence as depicted in **Figure 8**, a large number of them are controlled and provide real time data. For most of them the Snap4City mobile Apps also provide predictions and real time monitoring data.

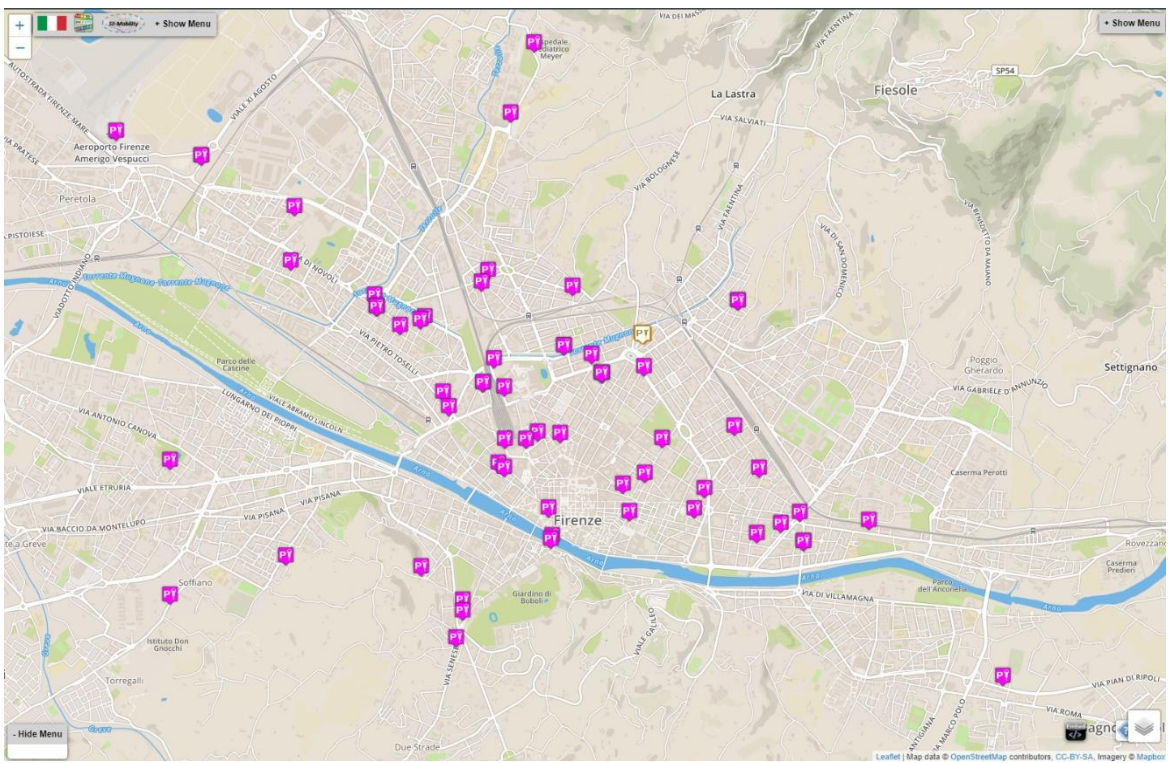

Figure 8: Main structures for parking facilities in Florence. Please note that only a part of them are monitored in terms of free parking lots in real time, https://servicemap.snap4city.org/

The trends of free parking lots have been already studied in the past with the aim of providing short and log terms predictions [16]. The typical trends for the days during working days and weekend where strongly different, resulting more critical in the working days in which a number of parking infrastructures run out of free parking lots. For example, in **Figure 9**, the comparison of the typical trends before the lockdown (left) and in the middle of the lockdown (right). Also in this case, there is some more information with respect to the Google global mobility report [13], [14] that registered a global reduction at the 30%. In effect also from our data the global average reduction has been at the 31%, while the detailed data provide much more meaning in understanding how this 30% is actually composed.

As reported in **Table 1**, the percentage of reduction in the exploitation of parking facilities in Florence is reported. The reduction has been estimated by clustering them according to their main purpose and services to: hospital, stations, social hubs, market, and justice. It is evident that the reduction of mobility at the 20% (above reported and discussion) also impacted on the parking lots serving the railway stations. Markets area obtained a lower reduction due to their primary necessity of services. A strong reduction has been registered to justice and social oriented parking area that have been activities totally stopped by the lockdown.

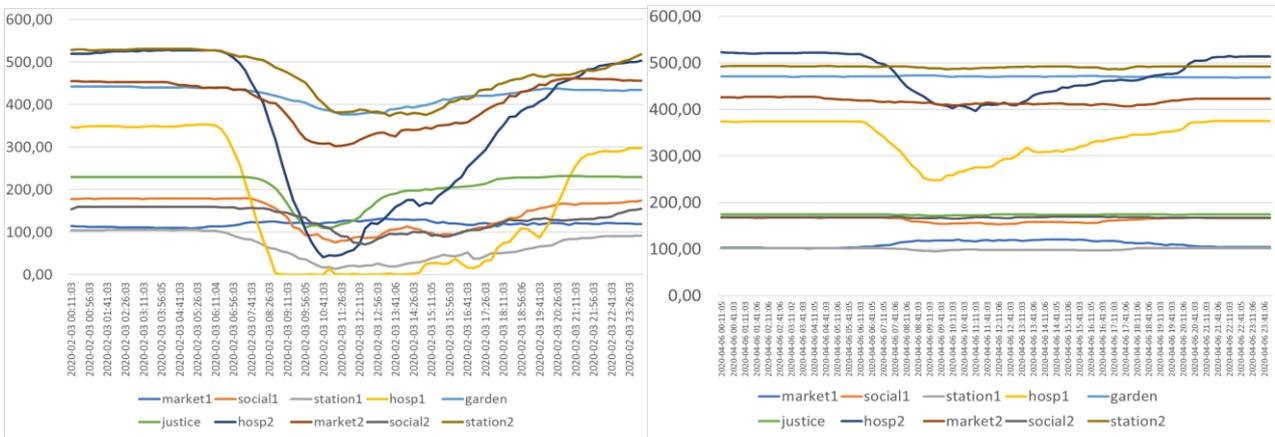

Figure 9: Typical Trends of the free parking lots on the first Monday of February (left) and April (right) for parking facilities in Florence.

| main Purpose | % of reduction |
|---|---|
| Hospitals | 55,9 |
| railway stations | 84,2 |
| social hubs | 85,8 |
| markets | 45,1 |
| justice | 96,1 |
| AVERAGE REDUCTION OF | 69,8 |

Table 1: Reduction of exploitation in parking facilities in Florence depending on their main utilization and purpose.

## 6. Impact of lockdown on environmental data and deductions

**Figure 10** reports the air quality and pollution station in the Florence area. They are a small part of the air quality and pollution stations of which data are collected by Snap4City. Their data come from sensors of CNR IBE, and ARPAT. Not each of them, but in total they are capable to read: PM10, PM2.5, CO, CO2, NO, NO2, O3, temperature, humidity, and few more.

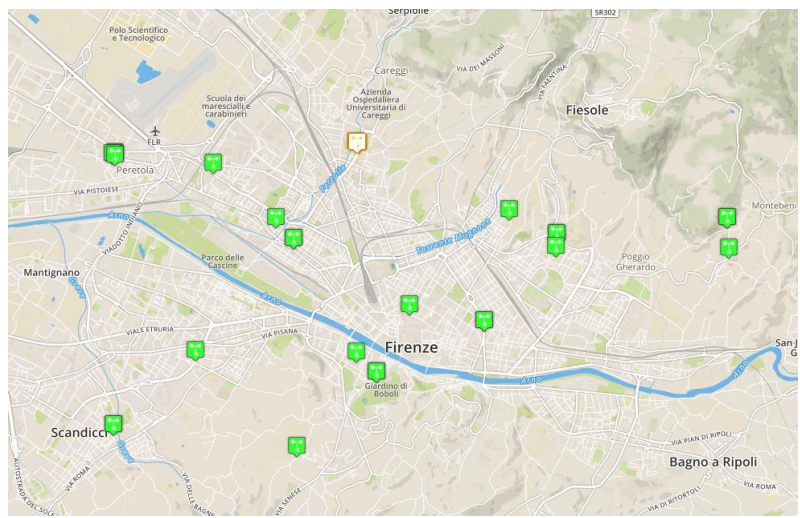

Figure 10: Air quality and pollution sensors in the area of Florence, as from https://www.snap4city.org .

In the context of environmental monitoring relevant data sources that could be affected by the above described situation are the values of the pollutants assessed by the above-mentioned sensors which are located in the city. The most relevant parameters that could be affected by the lockdown should be (some of these descriptions have been derived from those of Wikipedia):

- **NO2, NO** as NOx Nitrogen Oxides. NOX is a generic term for the nitrogen oxides that are most relevant for air pollution, produced from combustion, namely nitric oxide (NO) and nitrogen dioxide (NO2). NOx gases are also produced naturally by lightning. Other sources are house heating and industrial plants.
- **CO**: Carbon monoxide is a colorless, odorless, and tasteless gas that is slightly less dense than air. It is toxic to animals that use hemoglobin as an oxygen carrier (both invertebrate and vertebrate) when encountered in concentrations above about 35 ppm, although it is also produced in normal animal metabolism in low quantities, and is thought to have some normal biological functions. In the atmosphere, it is spatially variable and short lived, having a role in the formation of ground-level ozone.
- **CO2:** Carbon dioxide (chemical formula CO2) is a colorless gas with a density about 60% higher than that of dry air. It occurs naturally in Earth's atmosphere as a trace gas. The current concentration is about 0.04% (412 ppm) by volume, having risen from pre-industrial levels of 280 ppm. Natural sources include volcanoes, hot springs and geysers, and it is freed from carbonate rocks by dissolution in water and acids. Because carbon dioxide is soluble in water, it occurs naturally in groundwater, rivers and lakes, ice caps, glaciers and seawater. It is present in deposits of petroleum and natural gas. CO2 is produced by all aerobic organisms when they metabolize carbohydrates and lipids to produce energy by respiration. It is returned to water via the gills of fish and to the air via the lungs of air-breathing land animals, including humans.
- **PM10, PM2.5:** Particulate Matter measures are expressed in microgram for cube meter of particles of 10/2.5 micrometers or less, PM2.5 are more critical. The size of the particle is a determinant respiratory tract the particle will come to rest when inhaled. Larger particles are generally filtered in the nose, but particulate smaller than about 10 micrometers can settle in the bronchi and cause health problems. The 10-micrometer does not represent a strict boundary between respirable and non-respirable particles. The sources can be natural (soil erosion, marine spray, volcanoes, forest fires, pollen dispersion, etc.) or anthropogenic (industries, heating, vehicular traffic and combustion processes in general). The major components of atmospheric particulate matter are sulphate, nitrate, ammonia, sodium chloride, carbon, mineral dust and it is estimated that in some urban contexts more than 50% is of secondary origin.

A qualitative impact of lockdown can be observed comparing the February/March period with respect to April in the most crowded areas of *Gramsci* and *Ponte alle Mosse* in Florence. This view is reported **in Figure 11**, in which a Snap4City dashboard is reported. The different trends allow to compare the traffic flow (that has been strongly affected by the lockdown) with respect to the environmental variables such as: PM10, NO2, CO, etc. Among them the most influenced by the lockdown has been the NO2 due to its dependence to the traffic flow, heating and industrial activities. On the other hand, also the other pollutants have been influenced by the reduction of human activities.

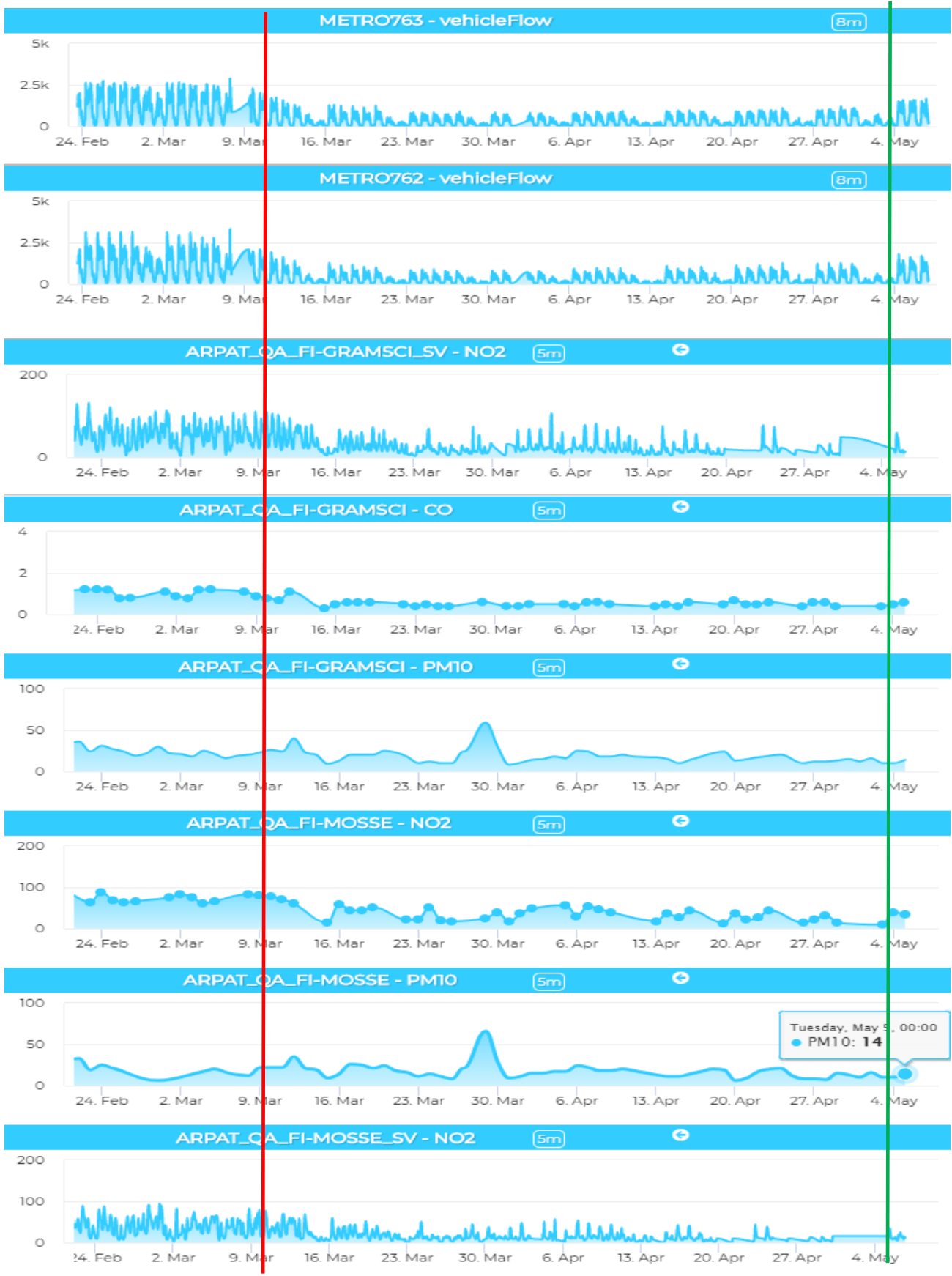

Figure 11: Comparison of the trends of different sensors values from February to May 2020. Red the beginning of the lockdown and red the reopening of a relevant part of the activities.

In order to assess the impact of the lock down on those environmental and air quality indicators a comparison of the average values of the above mentioned pollutant has been estimated to see the percentage of decrement with respect to the normal period, namely comparing February 2020 with respect to April (the central part of the lockdown). The results are reported in **Table 2**, from which is it is evident that the reduction of traffic at the 20% described above is one of responsible of the reduction of NO2 at the 39%. On the other hand, the reduction has not been proportional since the probably the heating of the houses have been increased in the same period, so that partially compensating the decrement of traffic and industry, which in Florence are not impacting very much in the city. Please note that the effects of the traffic into cities and the estimation of predictions for the NOX on the basis of traffic, wind direction, and structured of the city are studied and produced by the TRAFAIR project, of which the results for Tuscany area are available on Snap4City Dashboards [17].

|  | CO2 | CO | NO2 | PM10 |
|---|---|---|---|---|
| Percentage at which the average value of the pollutant has been reduced (February wrt April, 2020) | 41,06 | 47,54 | 38,84 | 61,99 |
| Percentage at which the MAX value of the pollutant has been reduced (February wrt April 2020) | 38,89 | 34,62 | 63,02 | 60,21 |

Table 2: Reduction of the pollutant in the period of Lockdown with respect to February 2020.

## 7. Conclusions

The analysis performed identified a strong reduction of the mobility and transport activities coherently to what has been described by Google Global mobility report and in addition putting in evidence a number of detailed aspects that could allow us to better judge the impact of the reopening on an eventual revamping of the infection. To this end, the collected data from the field have been compared from those of google and some considerations with respect to the Imperial college Report 20 have been derived. It is probably too early to derive conclusions since this report has been written the 6[th] of May, only 2 days after the formal reopening. The resulting traffic volume has been repristinated to more than the 50%, which is a +30% with respect to the lockdown situations. For parking, we still have to see an increment of their usage with respect to the lockdown conditions. As regard the environmental variable and thus or air quality the amount of pollutant has been strongly reduced by the lockdown. The largest reductions have been recorded for NO2 (also due to the reduction of traffic), CO, PM10 and CO2 for the reduction of human activities.